\documentclass[12pt,eqsecnum,nofootinbib,floats,article,aps,prd,floatfix,titlepage,superscriptaddress]{revtex4} 
\usepackage{epsfig}
\usepackage{graphicx}
\usepackage{bm}
\usepackage{amsmath}
\usepackage{amssymb}
\usepackage[multiple]{footmisc}

\usepackage{epstopdf}
\begin{document}

\title{Bubble Universe Dynamics After Free Passage}

\author{Pontus Ahlqvist}
\email{pontus@phys.columbia.edu}
\affiliation{Department of Physics, Columbia University, New York, New York 10027, USA}
\author{Kate Eckerle}
\email{ke2176@columbia.edu}
\affiliation{Department of Applied Mathematics, Columbia University, New York, New York 10027, USA}
\author{Brian Greene}
\email{greene@physics.columbia.edu}
\affiliation{Department of Physics and Department of Mathematics, Columbia University, New York, New York 10027, USA}

\begin{abstract}

We consider bubble collisions in single scalar field theories with multiple vacua. Recent work has argued that at sufficiently high impact velocities, collisions between such bubble vacua are governed by `free passage' dynamics in which field interactions can be ignored during the collision, providing a systematic process for populating local minima without quantum nucleation. We focus on the time period that follows the bubble collision and provide evidence that, for certain potentials, interactions can drive significant deviations from the free-passage bubble profile, thwarting the production of bubbles with different field values.

\end{abstract}

\maketitle
\section{Introduction}
Cosmological phase transitions, inflationary cosmology, as well as the putative landscape of string theory, all invoke various aspects of bubble universe dynamics. In this regard, the classic works \cite{Coleman,ColemanDeLuccia} play a key role in understanding the quantum nucleation of bubble universes with different scalar field expectation values. More recently, the works \cite{Easther; Giblin, Giblin} gave evidence that a distinct classical process, involving bubble collisions, provides an alternate--and efficient--mechanism for moving from one vacua to another. Central to the results of \cite{Easther; Giblin, Giblin} is the so-called `free passage' approximation in which self interactions of the scalar field are subdominant to other terms in the dynamical equations, and are thus ignored. 

In this note, we consider the epoch following a collision between two bubble universes to determine if the free passage field profile continues to guide the field dynamics and, more specifically, whether the field continues on a trajectory toward a new bubble vacua as suggested by the free passage solution. We find conditions under which we expect the free passage expectations to be violated, and confirm these expectations through numerical simulations.

\section{Background}

Recent works \cite{Easther; Giblin, Giblin, Aguirre-Johnson, Johnson} indicate that ultra-relativistic bubble collisions provide a mechanism for efficiently moving between vacua.  Generally speaking, an accurate description of the collision between two bubbles embedded in a parent false vacuum requires using the full nonlinear equations of motion. But the ultra-relativistic limit offers a great simplification, as the nonlinearities become subdominant \cite{Easther; Giblin}, and so the solution is given by superposing two single bubble solutions. This is the free passage approximation.

Qualitatively, free passage is accurate because in the large Lorentz factor limit, the kinetic energy dominates the potential up until and for short time after the collision. The reason is that before the collision, both spatial and time derivatives of the field in the walls are large, but $\partial V/\partial \phi\sim0$ everywhere. And as the walls become ever more Lorentz contracted the amount of time it takes for the walls to pass through each other diminishes, and $\partial V/\partial \phi$ is not large enough to produce an acceleration great enough to significantly alter the field's free evolution during the collision, and so the walls simply superimpose and pass through each other. 

More specifically, a collision is sufficiently relativistic for the free passage approximation to be valid if the Lorentz factor of the walls measured by an observer in the rest frame of the collision satisfies two inequalities \cite{Easther; Giblin, Giblin}. One of these comes from energy conservation, and involves the ratio of the heights of the barriers between the relevant vacua.  The second condition comes from ensuring that the walls make it past each other before deviations from the homogeneous solution have time to grow. This latter condition is formulated in terms of the slope of the potential at the kicked field value (the field value just after the bubble walls collide), and the rest width of the solitons. (In both of these minimum-Lorentz-factor-conditions, there are overall dissipation coefficients which have yet to be related to parameters in the theory. )

The authors of \cite{Easther; Giblin} claim that after free passage, the field in the collision region rolls to the minimum of the basin of attraction to which it was propelled via the free passage kick. If the kicked field value happens to be in the basin of attraction of a new lower energy density vacuum, then the field in the collision region rolls to the new local minimum, exhibiting a coherent transformation to a new expanding bubble vacuum. 

It is worth noting, however, that the specific models considered in \cite{Easther; Giblin, Giblin}
respect various non-generic symmetries which may be responsible, in part, for the clean transition to the new vacuum. Namely, the models considered involve a potential with three nearly equidistant minima, which ensures that the free passage collision between two ``middle" vacuum bubbles propels the field almost exactly to a new vacuum value. Moreover, the potential studied was (nearly) symmetric about the bubble vacuum. Which raises the natural question: What is the post-free-passage dynamics for a more generic potential? When the simplifications/symmetries noted are no longer present, does the field continue to roll to the local minimum of the basin of attraction it lands in via free passage? 

To study this, we follow the approach taken by the authors in \cite{Giblin}, and consider soliton-anti soliton collisions in 1+1D, with more generic potentials. We invoke the thin wall approximation, which is valid when the two neighboring minima between which a bubble wall interpolates are nearly degenerate. In this case, the initial solitons are expressed as

\begin{align}
&\Box\phi=-\frac{\partial V}{\partial\phi}\\
&\phi(0,\vec{x})=f(|\vec{x}|-R)\\
&\dot\phi(0,\vec{x})=0
\end{align}
where $f(r)$ is the soliton associated with the degenerate potential, and the initial bubble radius is dependent on the potential-- in particular three times the 1D soliton's action divided by the difference in energy densities of the bubble and parent vacuum. Hence, for large $R$ the collision of two bubbles (nucleated sufficiently far apart that their walls reach relativistic speeds before colliding) looks effectively like the collision of domain walls. So, the collision of a soliton and anti-soliton in 1+1D, each boosted to some constant relativistic speed, $u$, is a relevant problem to consider. The initial value problem is as follows
\begin{equation}
\begin{aligned}
&\frac{\partial^2\phi}{\partial t^2}-\frac{\partial^2\phi}{\partial x^2}=-\frac{\partial V}{\partial \phi}\\
&\lim_{t\rightarrow -\infty}\phi(t,x)=f(\gamma (x-ut))+f(-\gamma(x+ut))-\phi_A\\
&\lim_{t\rightarrow-\infty}\dot\phi(t,x)=-\gamma u\left(f'(\gamma (x-ut))+f'(\gamma (x-ut))\right)
\end{aligned}
\label{IVP}
\end{equation}
where $\gamma$ is the Lorentz factor, the potential, $V$, has degenerate minima $\phi_A$, and $\phi_B$, and the soliton $f(x)$ approaches $\phi_B$ as $x\rightarrow -\infty$, and $\phi_A$ as $x\rightarrow \infty$. To simplify notation in the following section, we now label the left moving and right moving solitons as follows,
\begin{align}
&f_{\rm R}(t,x)=f(\gamma(x-ut))\\
&f_{\rm L}(t,x)=f(-\gamma(x+ut))
\end{align}
and define the free passage solution, 
\begin{equation}
\phi_{\rm FP}(t,x)=f_{\rm R}(t,x)+f_{\rm L}(t,x)+\phi_A
\end{equation}
Lastly, we note the the field in the region in between the walls before the collision, which is in $\phi\sim\phi_A$, is shifted by the sum of the changes in field values across each of the walls, here simply $\delta\phi=2(\phi_B-\phi_A)$. This field deviation is the mathematical form of the free passage kick. 

Let's now turn to the post-free-passage field dynamics.

\label{sec:Background}
\section{Heuristic Argument}
\label{sec:heuristic}
To study the classical evolution of the system after the free passage kick, we write the field as
\begin{equation}
\phi(t,x)=\phi_{\rm FP}(t,x)+\sigma(t,x).
\end{equation}
Note that after the collision, $\phi_{\rm FP}$ takes on the value $2\phi_B-\phi_A$ within the collision region and $\phi_A$ outside of it, and so all the subsequent dynamics are encoded in $\sigma$. Substituting this form into the original equation of motion for $\phi$ yields
\begin{equation}
\Box\sigma=\left.\frac{\partial V}{\partial\phi}\right|_{\phi_{\rm FP}+\sigma}-\left.\frac{\partial V}{\partial\phi}\right|_{f_{\rm R}}-\left.\frac{\partial V}{\partial\phi}\right|_{f_{\rm L}}.
\end{equation}
Shortly after the collision $\sigma$ remains small and so we expand to obtain
\begin{equation}
\Box\sigma=\left(\left.\frac{\partial V}{\partial\phi}\right|_{\phi_{\rm FP}}-\left.\frac{\partial V}{\partial\phi}\right|_{f_{\rm R}}-\left.\frac{\partial V}{\partial\phi}\right|_{f_{\rm L}}\right)+\left.\frac{\partial^2 V}{\partial\phi^2}\right|_{\phi_{\rm FP}}\sigma+\mathcal{O}(\sigma^2).
\label{EOM-sigma}
\end{equation}
By dropping the term linear in $\sigma$ we find that the field in the collision region is driven by the term
\begin{equation}
\left.\frac{\partial V}{\partial\phi}\right|_{\phi_{\rm FP}}.
\end{equation}
Note that we have dropped the other two zeroth order terms appearing in equation (\ref{EOM-sigma}) since they evaluate to 
zero within the collision region. The field dynamics is then driven by the slope of the potential at $2\phi_B-\phi_A$. During the bubble collision, sufficiently high relative wall velocities ensure that the resulting field evolution will be much smaller than the free passage kick. After the bubble collision, the natural expectation is  that if the field falls in the basin of attraction of another vacuum, the field will subsequently roll to it. But this expectation relies on dropping the term linear in sigma, and while $\sigma$ may start out small, it can quickly grow\footnote{This is of course not true when $2\phi_B-\phi_A$ happens to be near a local minimum}. We will focus on cases in which a growing $\sigma$ can significantly alter the post collision evolution. Indeed, we will see that there are cases in which the term linear in $\sigma$ drives the field back toward $\phi_B$ thereby undoing the work of free passage. 

To motivate this result, and to assess its genericity, let's consider the relative strength of the two lowest order terms in $\sigma$.

\subsection{The First Order Term}

Here we isolate the effect of the term linear in $\sigma$ by dropping the zeroth order term in equation (\ref{EOM-sigma}) and considering the following equation of motion
\begin{equation}
-\ddot\sigma=-\frac{\partial^2\sigma}{\partial x^2}+\left.\frac{\partial^2 V}{\partial\phi^2}\right|_{\phi_{\rm FP}}\sigma.
\label{EOM-first-order-term-only}
\end{equation}
Our approach to analyzing the dynamics governed by \ref{EOM-first-order-term-only} is informed by Coleman's proof of the stability of solitons \cite{aspects-symmetry}; we identify the right hand side as an ordinary quantum mechanical Hamiltonian operator acting on $\sigma$, with an `analog' quantum potential given by
\begin{equation}
V_{\rm QM}(t,x)=\left.\frac{\partial^2 V}{\partial\phi^2}\right|_{\phi_{\rm FP}}.
\end{equation}

\begin{figure}[t]
\centerline{\includegraphics[width=.8\textwidth]{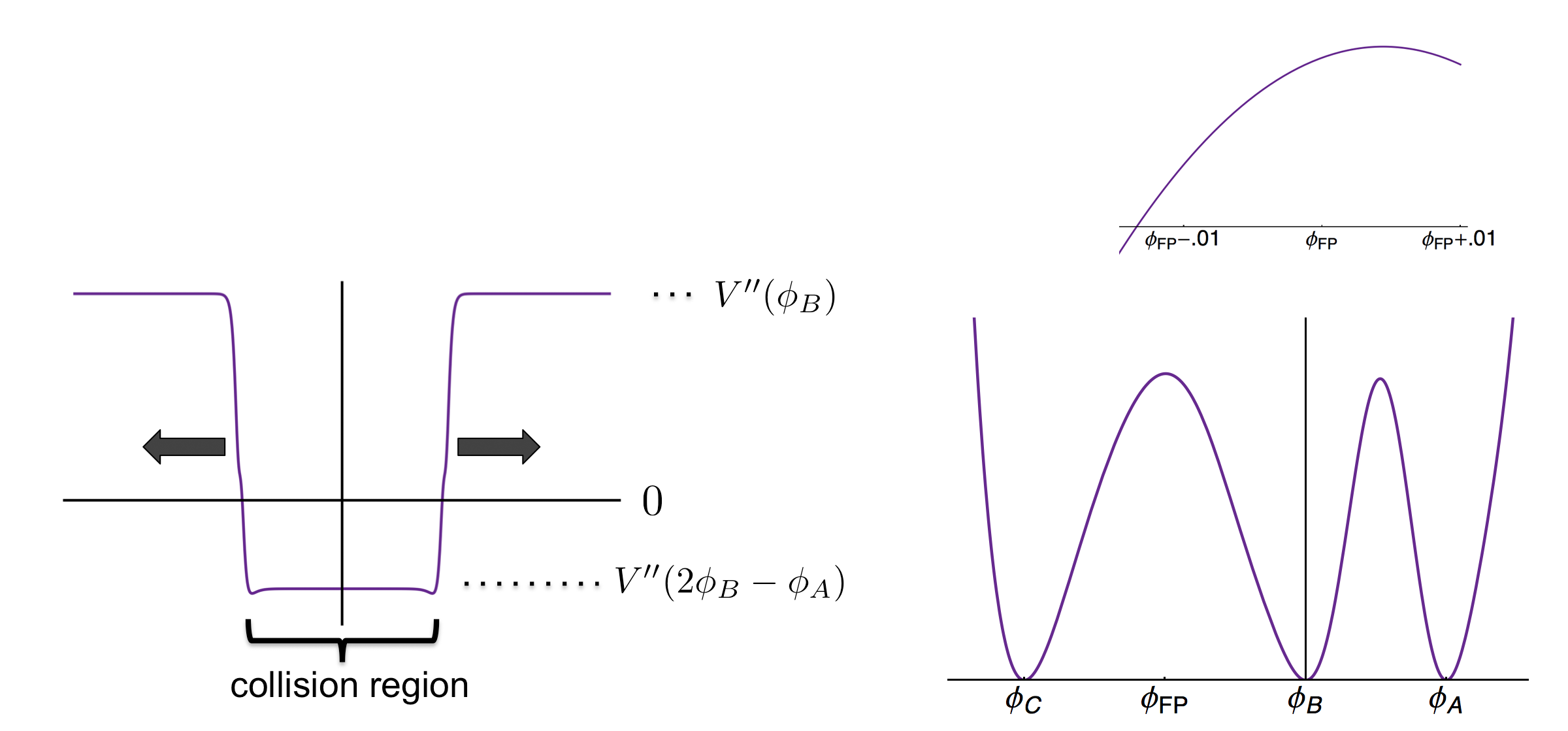}}
\caption{To the left is an example of an analog quantum potential after collision. This example is typical of those models that permit unstable mode(s) after collision. The particular collision this is associated with the one pictured in figures \ref{fig-unstableMode}, and \ref{fig-sigma}, whose potential is pictured to the right. We've included a magnified plot of the potential near the free passage field value on the upper right.}
\label{fig-Vqm}
\end{figure}

Qualitatively, this potential looks like a smoothed out finite square well (or barrier) with value $V''(\phi_B)$ outside the collision region, and $V''(2\phi_B-\phi_A)$ inside. In the relativistic limit, the walls become Lorentz contracted and the  potential looks increasingly like a widening square well (or barrier, depending on the magnitudes of $V''(2B-A)$ and $V''(B)$). An example of such an analog quantum potential is shown in figure \ref{fig-Vqm}. 

Of particular interest are cases where the field makes it into the basin of attraction of a new local minimum -- what we have been calling $\phi_C$ -- but does not migrate sufficiently far into the basin for the curvature of the potential to turn positive. In such cases, our quantum potential is a widening square well with a negative bottom, which means negative energy eigenstates are possible. While negative energies are not particularly special in the Schr\"odinger equation, here they are suggestive of exponentially growing (and decaying) modes. 

The instantaneous ground state of this system starts with energy $V''(\phi_B)>0$ immediately after the collision, and then decreases as the well widens, approaching $V''(2\phi_B-\phi_A)<0$ as $t\rightarrow \infty$. Consider a time $t_0>0$ when the ground state energy is negative. Had our Hamiltonian been time independent, the ground state, $\psi_0(t_0,x)$, would have evolved exponentially according to $\exp(\pm \sqrt{|E|}(t-t_0))$. While the time dependence of the system does factor in, we can still trust that a growing mode is present as long as the ground state changes sufficiently slowly\footnote{More precisely, as long as the overlap between two successive ground states is sufficiently close to $1$ so that the exponentially growing factor dominates it, the mode will grow. There is always a $t_0$ large enough such that this is the case. For a more thorough treatment consult the Appendix.}. Note that implicitly we assume the walls are boosted to a sufficiently large speed that linearization is still valid at $t_0$.

Of course, at time $t_0$ the ground state component of the deviation from free passage may be positive or negative. So the mode's contribution to the field's evolution post free passage can be towards or away from $\phi_C$. To clarify this we write
\begin{equation}
\sigma(t_0+dt,x)\sim [\alpha_0 \exp(\sqrt{|E_0|}dt)+\beta_0 \exp(-\sqrt{|E_0|}dt)]\psi_0(t_0,x) + ...
\end{equation}
where ``..." represents the contributions from the remaining modes. These consist of scattering states, which are all stable, and additional bound states which, though possibly unstable, are less unstable than the ground state. Hence we expect dynamics (given by the EOM for $\sigma$ in this subsection, obtained from keeping only the first order forcing) to be dominated by the ground state. The ground state, like any state, is not unique up to a phase. Let us take $\psi_0$ to be real and positive. Then $\alpha_0$ is real, and the ground state contributes a push toward $\phi_C$ if the sign of $\alpha_0$ is the same as $\phi_C-(2\phi_B-\phi_A)$ (negative for us), and vice-versa (i.e. retreat to the original bubble vacuum). 

In terms of the deviation's initial conditions (given at time $t_0$ that satisfies the above conditions), $\alpha_0$ is
\begin{equation}
\alpha_0= \int \psi_0(t_0,x)[\sigma(t_0,x)+\frac{1}{\sqrt{|E_0|}}\dot\sigma(t_0,x)].
\end{equation}
The initial conditions for $\sigma$ are essentially the accumulated effect of the full zeroth order forcing term on $\phi$ up until $t_0$ (ignoring $\mathcal{O}(\epsilon)$ terms in $\sigma$'s EOM is valid until $t_0$). Recall this term is given by,
\begin{equation}
-\frac{\partial V}{\partial\phi}\bigg{|}_{\phi_{FP}(t,x)}+\frac{\partial V}{\partial\phi}\bigg{|}_{f_L(t,x)}+\frac{\partial V}{\partial\phi}\bigg{|}_{f_R(t,x)}.
\label{forcing-zeroth}
\end{equation}

Here we demonstrate that our analysis reproduces the correct behavior in the limit $u\rightarrow1$, where $\phi_C$ bubble nucleation is successful. The deviation the instant after collision, which results from integrating the above forcing against the appropriate Green's function, vanishes in this limit because it is suppressed by $u/\gamma$ (see eq. 16 in \cite{Giblin}). After this instant, the only non-negligible forcing in the collision region is simply $-V'(2\phi_B-\phi_A)$. Consequently, $\alpha_0$ will have the same sign as $\phi_C-(2\phi_B-\phi_A)$, since $2\phi_B-\phi_A$ is in the basin of attraction of $\phi_C$. Hence, both zeroth and first order forcing terms in $\sigma$'s EOM would result in the field being pushed toward $\phi_C$ if treated independently. Also, note that qualitatively our analysis reproduces the correct spatial dependence of the solution post collision-- the distance the field rolls toward $\phi_C$ is peaked in the center of the collision region and decreases to nearly zero at the walls. So long as the contribution to $\sigma(t_0,x)$ and $\dot\sigma(t_0,x)$ from the forcing \emph{after} the walls finish passing through each other dominates the contribution accumulated until this time, we expect successful bubble $\phi_C$ bubble nucleation. 

If, on the other hand, the speed is not sufficiently large that the contribution from the forcing before/during collision is negligible it is possible for $\alpha_0$ to be of the opposite sign as $-V'(2\phi_B-\phi_A)$. In such a case there is a competition between the zeroth and first order forcing terms. This raises a perhaps surprising prospect. The field could realize enough of the free passage kick that the collision region is indeed taken into the basin of attraction of $\phi_C$, but fall short enough that $\alpha_0$ is sufficiently large in magnitude that the first order forcing term dominates the dynamics. This would mean that, despite having made it into the basin of attraction of a new vacuum via free passage, the field in the collision region would nonetheless be pulled back uphill towards the original bubble vacuum. If the effect is significant enough the field would make it all the way back over the barrier, into the old bubble's basin of attraction thereby undoing the kick from free passage and preventing a bubble of new vacuum from forming. In particular, this would mean that consideration of the post-collision dynamics raises the minimum collision speed necessary to complete the free passage transition.

\subsection{Comparison of zeroth and first order terms' time scales}

Whether the instability identified in the previous section is realized depends on how the time scale associated with its growth compares with that associated with the zeroth order term. If the slope at $2\phi_{B}-\phi_A$ is sufficiently large, the field in the collision region accelerates quickly, and the window for exciting the growing mode is lost. The time scale associated with the zeroth order term can be determined by dropping the Laplacian of the deviation and explicitly solving the resulting differential equation,
\begin{equation}
\ddot\sigma\thickapprox V'_{\rm 2B-A} \rightarrow \sigma\sim \frac{V'_{\rm 2B-A}}{2}t^2+\mathcal{O}(t).
\end{equation}
Hence, the time scale associated with the zeroth order driving term is 
\begin{equation}
t_{\mbox{zeroth}} \sim \frac{1}{\sqrt{V'_{\rm 2B-A}}}.
\end{equation}
The time scale associated with the first order term is roughly given by the time for one e-folding, which in turn depends on the depth of the analog quantum potential,
\begin{equation}
t_{\mbox{linear}}\sim\frac{1}{\sqrt{|V_{2B-A}''|}}
\end{equation}
This suggests that in models with 
\begin{equation}
\sqrt{\frac{|V''_{2B-A}|}{V'_{2B-A}}}\gg 1,
\label{condition}
\end{equation}
the field in the collision region may not simply roll down towards $\phi_C$, but rather be significantly influenced by the unstable mode. As indicated previously, whether the mode grows to drive the field toward $\phi_B$ or $\phi_C$ depends on the initial conditions for the deviation.

In the following section we undertake a detailed numerical study of this issue and try to determine the threshold value for the expression in equation (\ref{condition}) for which the field returns to $\phi_B$ after the collision.

\section{Numerical Survey}
In the previous section we presented a heuristic argument for why the time evolution after a free passage kick takes the field into the basin of attraction of new vacuum ($\phi_C$) may bring the field back to the original vacuum ($\phi_B$), rather than causing it to roll to $\phi_C$. Again, even in such cases
 there exists a speed above which the ``naive" picture of dynamics-- free passage followed by evolution according to $-V'$ at the kicked field value-- will be realized. Our point, though, is that there can be potentials in which this threshold speed is greater than one would naively expect, due to the instability we've identified. For potentials in which $2\phi_B-\phi_A$ lies in the basin of attraction of a new vacuum, the naive expectation is that the threshold will have been passed if the kicked field in the collision region is within a small enough distance of $2\phi_B-\phi_A$ that it lies in the new minimum's basin of attraction. 

We want to study how the detailed features of the potential, such as the relative size (and sign) of the
 first and second derivatives at $2\phi_B-\phi_A$, determine whether the free passage dynamics nucleates a bubble of new vacuum, $\phi_C$.
Thus, we numerically simulate relativistic soliton-anti soliton collisions in models where $V'(2\phi_B-\phi_A)$, and $V''(2\phi_B-\phi_A)$ can be tuned.  In particular, we studied the following two potentials, each with degenerate local minima at $\phi = -2, 0, \mbox{and } 1$:
\begin{equation}
V(\phi)=\phi^2(\phi-1)^2(\phi+2)^2(\left(1+k_1\exp(-k_2(\phi+.2)^2)+k_3\exp(-4(\phi-.5)^2)\right)
\label{potential1}
\end{equation}
and,
\begin{equation}
V(\phi)=\phi^2(\phi-1)^2(\phi+2)^2\left(1+k_1\exp(-k_2(\phi+.2)^2)(\phi+k_4)+k_3\exp(-4(\phi-.5)^2)\right).
\label{potential2}
\end{equation}

The same initial value problem laid out in section \ref{sec:Background} is solved with the Cactus Computational Toolkit utilizing a fourth order Runge-Kutta method. For each choice of parameters $\{k_i\}$ we have determined whether the field in the collision region, after receiving its free passage kick, rolls toward the new vacuum at $\phi_C=-2$, or retreats to the original bubble vacuum $\phi_B=0$. The figure below displays the results. Runs are plotted in the $V'(2\phi_B-\phi_A)$-$V''(2\phi_B-\phi_A)$ plane, and the color indicates the outcome: purple for those that retreat back toward $\phi_B = 0$, and black for those that roll toward $\phi_C = -2$. 

\begin{figure}
\includegraphics{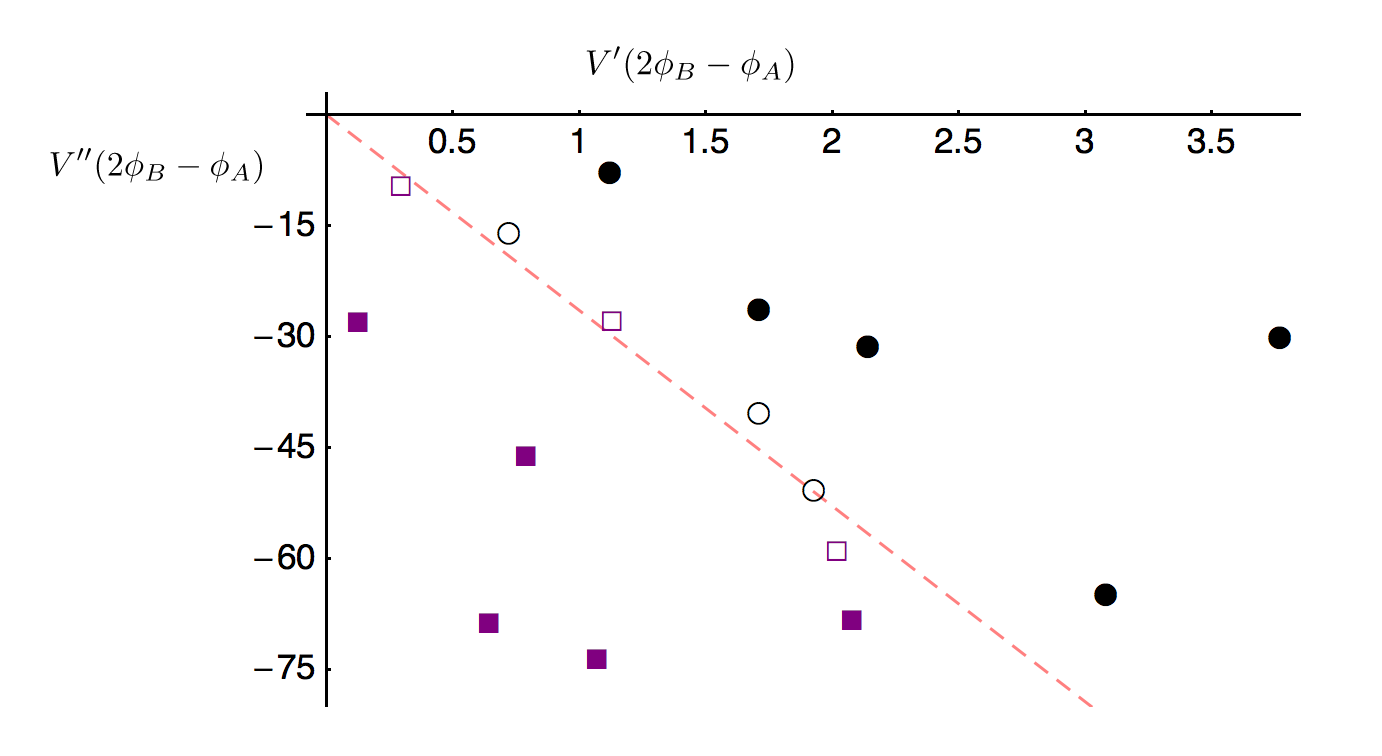}
\caption{The results of numerical simulations for various values of $V'(2\phi_B-\phi_A)$ and $V''(2\phi_B-\phi_A)$. Cases where the field retreats back toward $\phi_B$ are indicated by a purple square while those that continue toward $\phi_C$ are indicated by black circles. Note that there is a rough boundary that separates the two classes. We've unshaded the simulations deemed to lie along this boundary, and plotted a least squares fit to these points which has slope -26.5.}
\label{fig-numericalResults}
\end{figure}
The runs naturally separate based on the relative magnitude of $V''(2\phi_B-\phi_A)$, and $V'(2\phi_B-\phi_A)$, as expected. The boundary between the two regions is approximately linear, with slope $\sim -26.5$. We thus find that the threshold for successful new bubble nucleation via free passage may be increased above the level naively expected (that which lands the field in the basin of attraction of the new vacuum) for models with 
\begin{equation}
\frac{|V''(2\phi_B-\phi_A)|}{V'(2\phi_B-\phi_A)}\gtrsim 26.5.
\label{numericalCondition}
\end{equation}
Below we include snap shots of a collision in a model representative of those where the field in the collision region is pulled back into the basin of attraction of $\phi_B$. We do this to illustrate that the mechanism for this ``failure to nucleate'' is indeed our unstable mode. Notice that the retreat begins in the center of the collision region and eventually drags the rest of the interior back to $\phi_B$. For this particular simulation we used the potential given by \ref{potential1}, with parameters $k_1=1.85$, $k_2=1.54$, and $k_3=15$, along with a wall speed $u=.999$.
\begin{figure}[t]
\centerline{\includegraphics[width=1.2\textwidth]{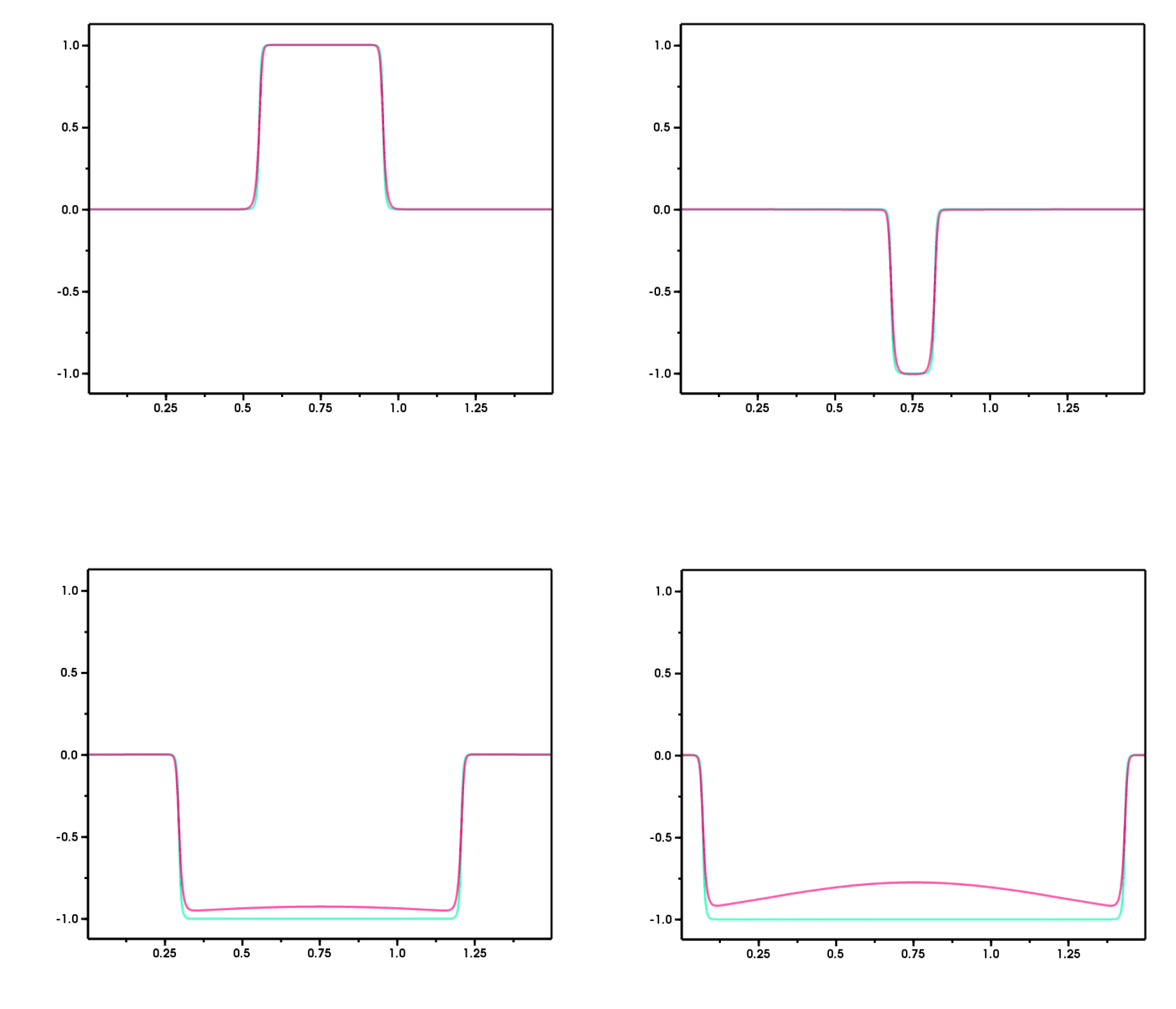}}
\caption{Here we present a representative case where the field retreats back to $\phi_B$ despite temporarily migrating into basin of attraction of the minimum at $-2$ after the collision. The solution to the field's EOM, in red, is plotted over the free passage solution, in green. (Note: Solitons in the field's initial conditions were approximated by \ref{approx-soliton}, whereas those used in the free passage solution were constructed by numerical inversion as discussed in the footnote on page \pageref{true-soliton}. As noted, the effect of using the more exact numerical approach is minimal.)}
\label{fig-unstableMode}
\end{figure}

\begin{figure}[t]
\centerline{\includegraphics[width=1.2\textwidth]{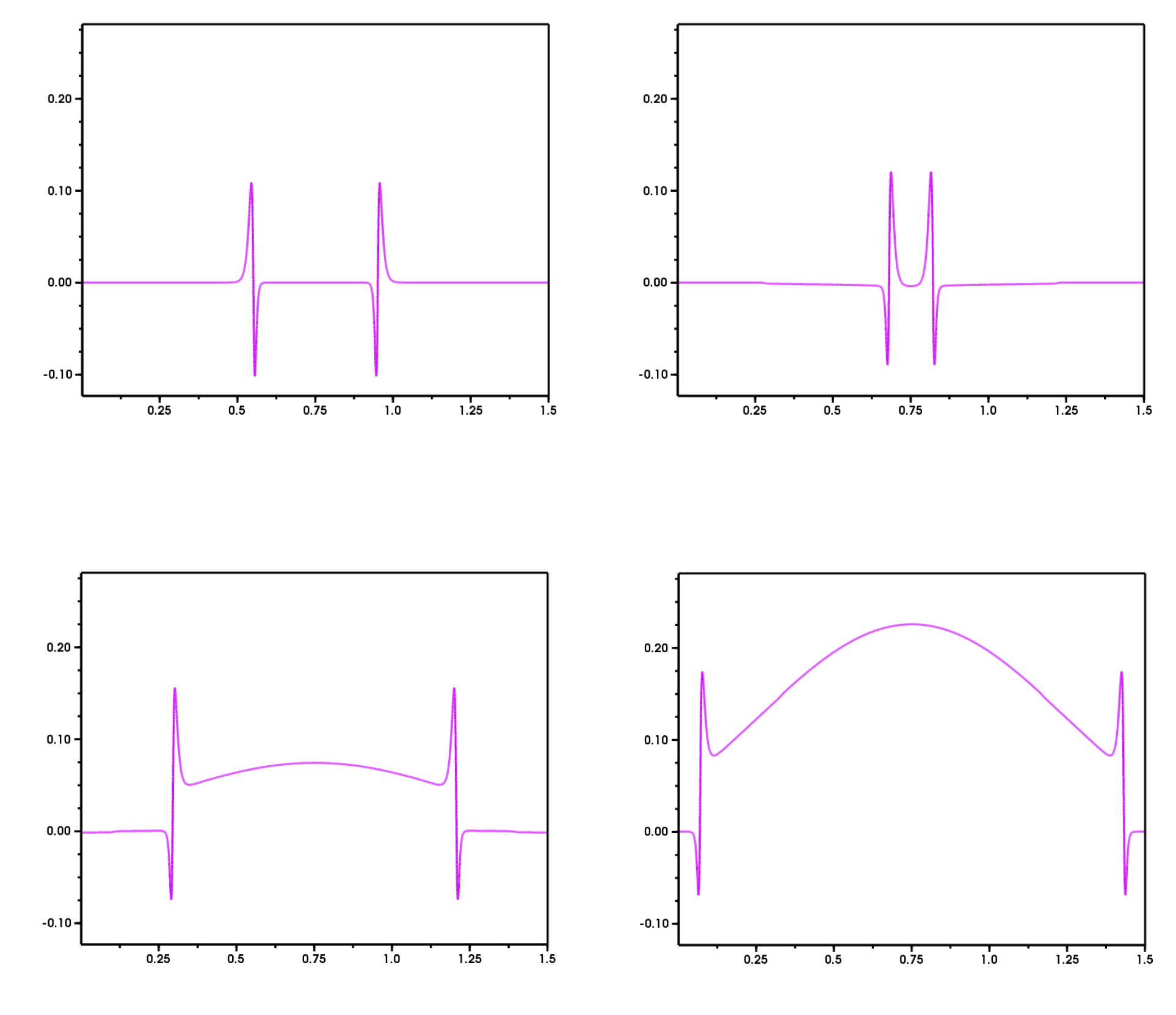}}
\caption{For each snap shot in figure \ref{fig-unstableMode} we plot the corresponding deviation from free passage, merely computed as the difference between the actual field solution and the free passage solution. Notice that the shape of the instability is similar to the ground state of a finite square well thus supporting our heuristic argument in section \ref{sec:heuristic}. The kinks at the wall locations indicate that the soliton approximation, \ref{approx-soliton}, produces a slightly wider soliton than the true one. The persistence of the kinks throughout the snapshots means that the approximate walls stay intact. Thus, retreat to the original bubble vacuum does not result from collapse of the walls, but rather from evolution of the field inside the collision region. The former behavior would not be an example of our effect since the ground state is nearly zero at the walls. Instead, it would be the ``temporary excursion" observed by \cite{Easther; Giblin, Giblin} at insufficiently relativistic speeds.}
\label{fig-sigma}
\end{figure}

\begin{figure}[htb]
\centerline{\includegraphics[width=1.2\textwidth]{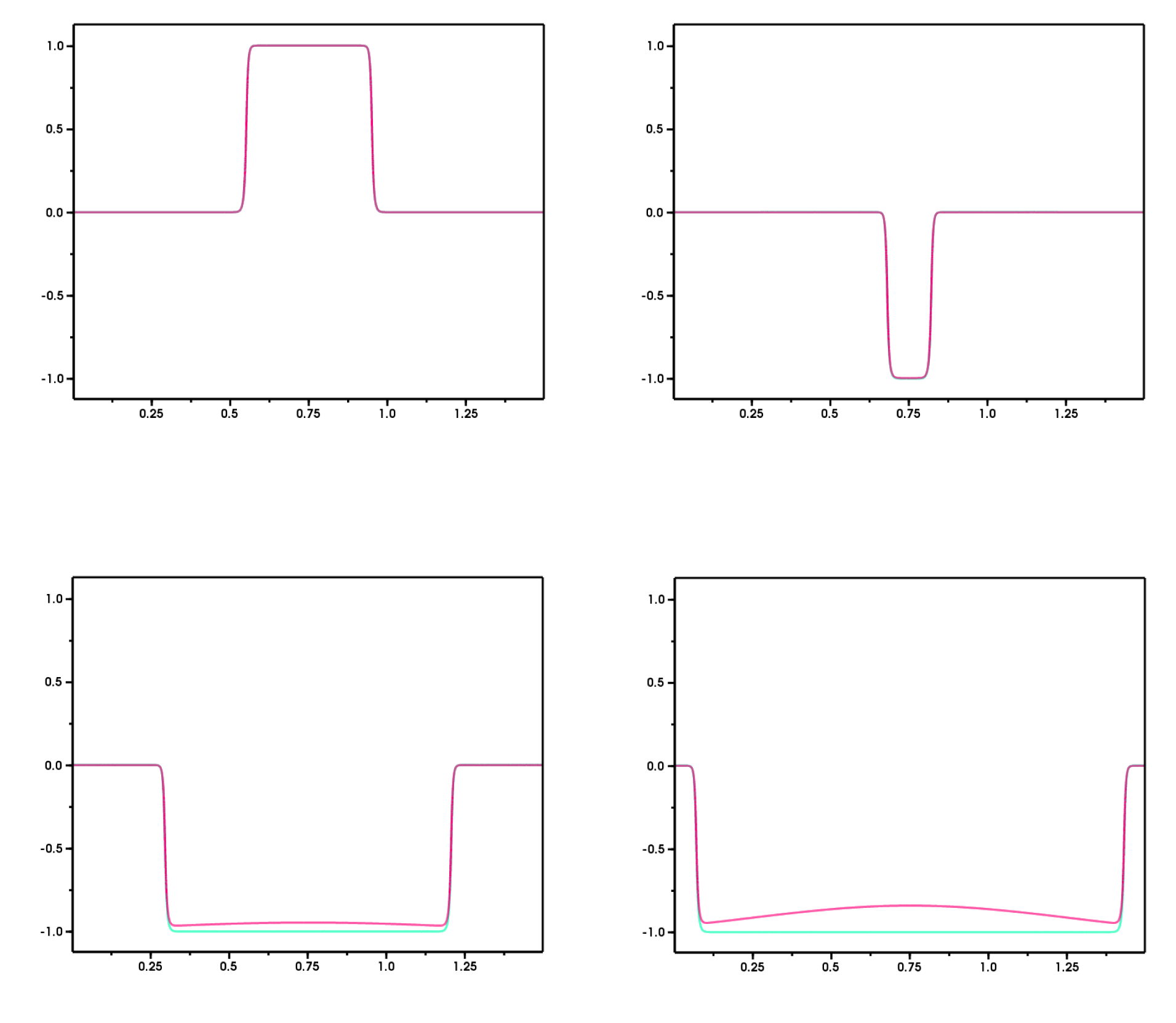}}
\caption{Results of the same collision pictured in figure \ref{fig-unstableMode}, only with solitons in the field's initial conditions constructed using the more accurate method described in the footnote on \pageref{true-soliton}. All parameter values, including initial wall locations, and speed, were identical for the two runs. The solution to the field's EOM is in red, and the free passage solution is in green.}
\label{fig-field-true}
\end{figure}

\begin{figure}[htb]
\centerline{\includegraphics[width=1.2\textwidth]{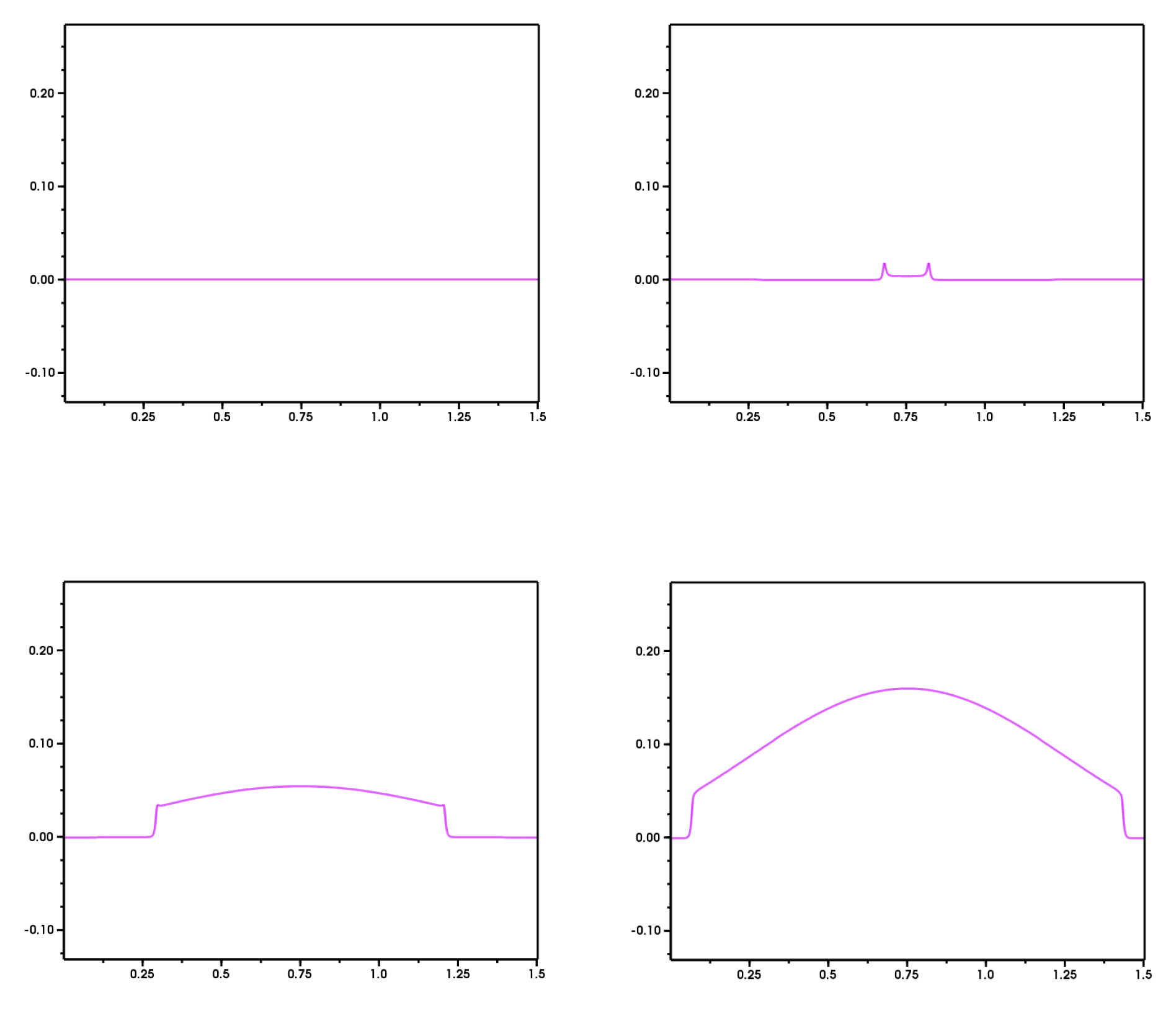}}
\caption{The deviation from free passage for the snap shots in figure \ref{fig-field-true}, computed as the difference between the actual field solution and free passage solution.}
\label{fig-sigma-true}
\end{figure}

For a given model, the soliton used to construct initial conditions in \ref{IVP} was approximated as follows:
\begin{equation}
f(x)=\frac{1}{\sqrt{1+3\exp(-x\sqrt{V''(1)})}}
\label{approx-soliton}
\end{equation}
This is a modification of the exact soliton that interpolates between the same vacua, $\phi = 0$ and $\phi = 1$, for the potential, 
\begin{equation}
V(\phi)=\phi^2(\phi^2-1)^2
\end{equation}
We chose the coefficient in the exponential to be $\sqrt{V''_{\phi =1}}$ in order to send any waves that developed from relaxation of the walls away from the collision region. The effect of the approximation is minimal, as can be seen by comparing the plots in figures \ref{fig-unstableMode}, and \ref{fig-sigma} with the corresponding ones obtained for the same collision simulated using the ``true" soliton in the field's initial conditions, pictured in figures \ref{fig-field-true}, and \ref{fig-sigma-true}.
\footnote{We have recently developed an approach for more accurately constructing initial conditions. Instead of using the closed form approximate expression given in \ref{approx-soliton}, we use a discrete approximation to the ``true" soliton obtained by numerical inversion of
\begin{equation}
x=\int_{f(0)}^{f(x)}\frac{d\phi}{\sqrt{2 V(\phi)}}.
\label{true-soliton}
\end{equation}
By limiting the values of the upper bound to the interval between two neighboring vacua the above is a definition for soliton $f(x)$ equivalent to the usual definition given in terms of a BVP. We have checked that this more precise form has only a marginal affect on our numerical survey of the soliton collisions.}

\section{Generality of instability}
While we have demonstrated the existence of an instability, it is now important to address the issue of generality. In particular, how finely tuned does a potential have to be to exhibit the kind of first and second derivatives required by equation (\ref{numericalCondition})? More precisely we must have this inequality satisfied at $2\phi_B-\phi_A$. 

Consider the slope of the potential evaluated very close to $2\phi_B-\phi_A$. As long as the potential is well behaved near that point and derivatives of order higher than two are not very large themselves, we may expand and write
\begin{equation}
V'(2\phi_B-\phi_A + \epsilon) \approx V'(2\phi_B-\phi_A) +  V''(2\phi_B-\phi_A)\, \epsilon.
\end{equation}

Since the second derivative is so much larger in magnitude than the first derivative, we can set the right hand side of this expression equal to zero by choosing a very small $\epsilon \sim 1/26.5$. In other words, we must be very close to an extremum of the potential. This places severe limitations on the types of potentials in which migration into the basin of attraction via free passage is only temporary. In particular, the locations of the minima of the potential must be so finely tuned that $2\phi_B-\phi_A$ is nearly at a maximum of the potential. 

Though our analysis is based on a first order taylor expansion of $V'$ about the free passage field value being valid for at least some $\delta>\epsilon\sim 1/26.5$, the results suggest it might be worth considering potentials where the condition on the ratio of the first and second derivatives is satisfied in a large interval around $2\phi_B-\phi_A$, albeit without requiring such a $\delta$. For instance, if one inserted an exponential segment, $V\sim\exp[-K \phi]$, into the standard three minima potentials we've considered thus far in an interval around the free passage field value, then the ratio of first to second derivatives everywhere in the interval will be $-1/K$. The size of the interval is arbitrary, so the distance between the locations of the parent and bubble vacuum can be moved liberally, without changing the ratio of the derivative at the kicked field value. 

Qualitatively, such a potential will have a plateau leading to a very steep cliff in between the bubble and new vacua, since $K$ must be chosen quite large. Preliminary numerical results indicate that uphill retreat post collision via our unstable ground state may in fact be realized for some potentials of this sort. To avoid a potential that is defined piece-wise, we ``carve out" two Gaussians from the plateau in the following way:
\begin{equation}
V(\phi)=1-\exp(-K(\phi-\eta))-\exp(-k_1(\phi-\phi_A)^2)-\exp(-k_2(\phi-\phi_B)^2)
\label{exponential-potential}
\end{equation}

\begin{figure}[t]
\centerline{\includegraphics[width=1\textwidth]{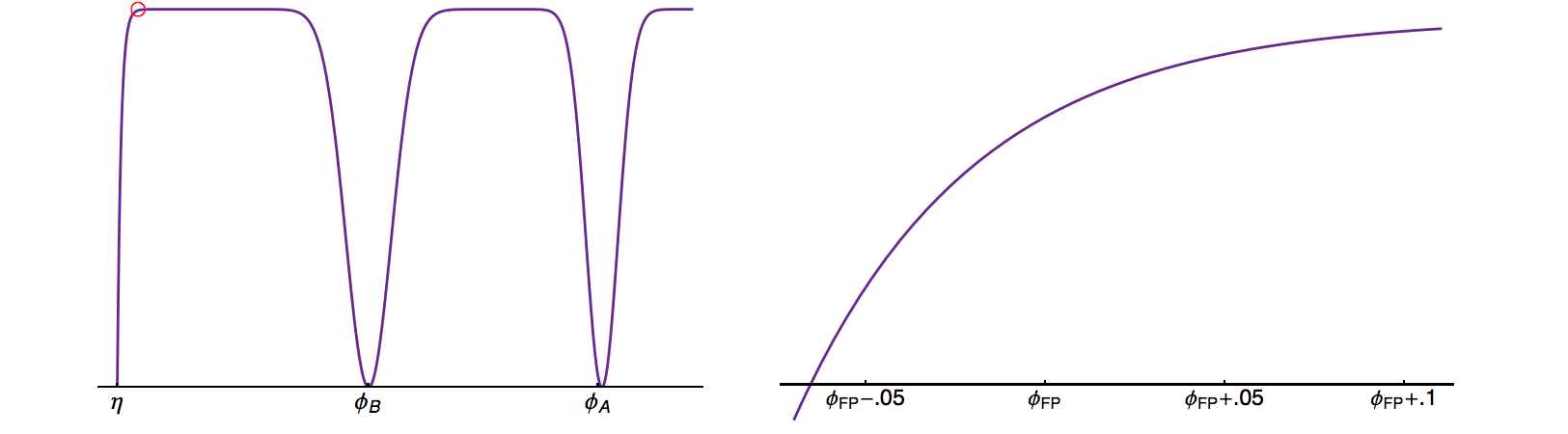}}
\caption{To the left is a plot of the potential given by \ref{exponential-potential} with parameter values $\eta=-3.5$, $K=20$,  $k_1=10$, $k_2=5$, $\phi_A=3.25$ and $\phi_B=0$. On the right we include a magnified plot of the potential near the free passage field location, $2\phi_B-\phi_A$ (circled in red on the unmagnified plot). The results of a soliton-anti soliton collision for this model are pictured in figure \ref{fig-expCollision}.}
\label{fig-expPotential}
\end{figure}

Once again, soliton-anti soliton collisions were simulated with $\phi_B$ as the bubble vacuum, and $\phi_A$ as the parent. We took $\eta=-3.5$, $K=20$,  $k_1=10$, $k_2=5$, $\phi_B=0$, wall speed $u=.99$, and varied $\phi_A$. We started with $\phi_A=3$, and observed retreat toward the $\phi_B$ vacuum. In each successive run, $\phi_A$ was increased. This has the effect, essentially, of leaving the potential in between $\eta$, and $\phi_B$ unchanged. As far as a collision is concerned, moving $\phi_A$ toward $3.5$ has the sole effect of moving the free passage kicked field value leftward along the plateau, ever closer to the cliff, without changing what the potential looks like between the cliff and the entrance to the basin of attraction of the $\phi_B$ vacuum. A plot of the potential for an example one of these simulations, that with $\phi_A=3.25$, can be found in figure \ref{fig-expPotential}.

\begin{figure}[t]
\centerline{\includegraphics[width=.9\textwidth]{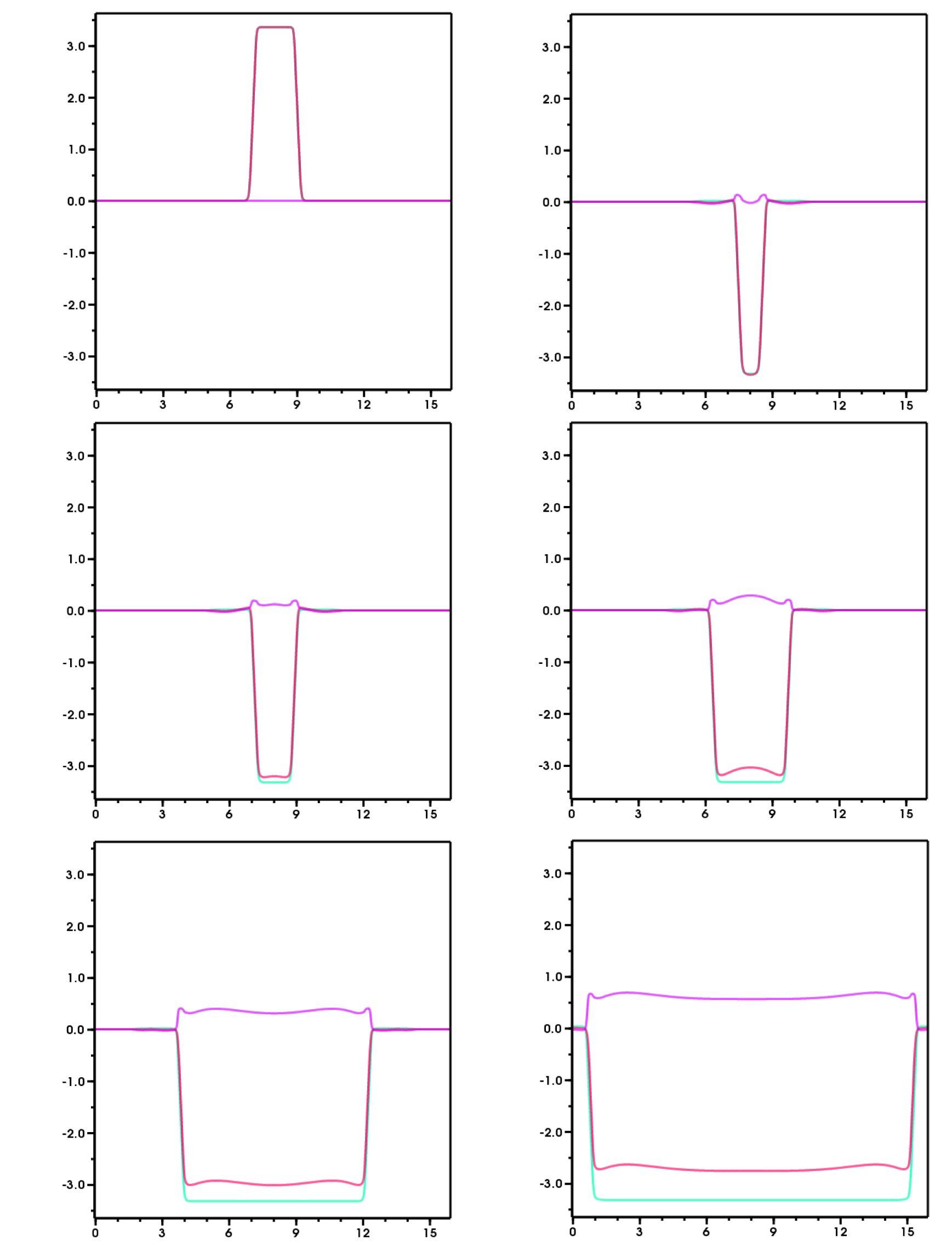}}
\caption{Above are snap shots of the collision of two $\phi_B$ bubbles nucleated in parent $\phi_A$ for the potential pictured in figure \ref{fig-expPotential}. The field is plotted in red, the free passage solution in green, and the deviation in purple. The field in the collision region is kicked via free passage to a location where $-V'$ is negative, yet nonetheless retreats in the positive direction-- running through an entire interval where $-V'$ is negative-- ultimately back to the original bubble's basin of attraction.}
\label{fig-expCollision}
\end{figure}

Our goal was to determine how close to the cliff the free passage kicked field could be (i.e. how large the magnitude $V'$ could be), and yet still exhibit this retreat (in the opposite direction of $-V'$) to the original bubble vacuum. Essentially, at what point do dynamics after collision switch from retreat to rolling off the cliff? We've found that this threshold $\phi_A$ value lies in between $3.25$ and $3.35$, as all simulations with $\phi_A\leq3.25$ retreated, and that with $\phi_A=3.35$ fell off. Snap shots of an example of a simulation in which retreat occurs, that with $\phi_A=3.25$, are displayed in figure \ref{fig-expCollision}.

Of course, the threshold $\phi_A$ value could be resolved further. These preliminary results nonetheless suggest that the somewhat surprising behavior suggested by our heuristic argument (based on linear analysis), observed in our corresponding numerical survey, and deemed finely tuned for the usual potentials considered (resembling those of simply polynomial form), may be quite general in other classes of models. This is because the effect might extend to models where the condition \ref{numericalCondition} is satisfied in a large (non negligible) neighborhood around the free passage kicked field location (hence not finely tuned), but where linearization about the free passage solution in the collision region is not valid in throughout the larger interval. 

\section{Discussion}

In this note, we've investigated bubble collisions in single scalar field theories whose potentials admit multiple local minima. Motivated by the findings of \cite{Easther; Giblin}, in which high velocity bubble collisions were shown to be governed by free passage dynamics, we've considered post collision evolution to determine the efficacy of classical nucleation of new bubble vacua. Specifically, we've used analytical and numerical arguments to assess the post-collision deviation from free passage dynamics, and identified potentials for which such deviations both rapidly grow and drive the field away from producing bubbles of with new vacuum field values. An interesting question touched on
here but deserving of more detailed study is the genericity of such deviations. 
A natural next step, with an eye toward applications to eternal inflation and the string landscape, is to consider these questions in multi-field models, a subject to which we shall return in \cite{massless-dof, curved-field-space}

\section*{Acknowledgements}
We thank Eugene Lim and John T. Giblin for helpful insights. This work was supported by the U.S. Department of Energy under grant DE-FG02-92ER40699.

\appendix*
\section{}

Here we discuss in more detail the claim that negative eigenvalues suggest the existence of unstable modes. We also illustrate that initial conditions for $\sigma$ can indeed yield exponential growth toward the $\phi_B$ vacuum, despite the field $\phi$ being in the basin of attraction of $\phi_C$. Once again we note that at some time $t_0$ after the collision, the deviation $\sigma(t_0,x)$, and its derivative $\dot\sigma(t_0,x)$ can be expanded in the instantaneous energy eigenbasis. While it would be correct to attach the time dependence $\exp(\pm i\sqrt{E}t)$, which are exponentially growing and decaying modes for negative $E$, to a time independent eigenstate, here it is not valid because the eigenstates change as the well widens. Consider instead discretizing the evolution. A time $dt$ later $\sigma$ and $\dot\sigma$ are stepped forward as follows: 
\begin{align}
\sigma(t_0+dt,x)&= [\alpha_0 \exp(\sqrt{|E_0|}dt)+\beta_0 \exp(-\sqrt{|E_0|}dt)]\psi_0(t_0,x)+ ...\\
\dot\sigma(t_0+dt,x)&= \sqrt{|E_0|}[\alpha_0 \exp(\sqrt{|E_0|}dt)-\beta_0 \exp(-\sqrt{|E_0|}dt)]\psi_0(t_0,x)+ ...
\end{align}
Of course at $t+dt$ we do not know how $\psi_0(t_0,x)$ evolves, since it is no longer an eigenstate of the Hamiltonian. Rather it is some linear combination of the new eigenstates. However, we can extract its new ground state component, evolve its contribution to $\sigma$ and $\dot\sigma$ in a similar manner. We absorb the exponentially decaying ground state mode, and all remaining evolved modes in ``..."
\begin{align}
\sigma(t_0+2dt,x)&=\alpha_0 \exp(\sqrt{|E_0|}dt)\exp(\sqrt{|E_0|}dt) \vec{\psi_0}(t_0+dt)\cdot\vec{\psi_0}(t_0) \psi_0(t_0+dt,x)+...\\
\dot\sigma(t_0+2dt,x)&=\sqrt{|E_0|}\alpha_0 \exp(\sqrt{|E_0|}dt)\exp(\sqrt{|E_0|}dt)] \vec{\psi_0}(t_0+dt)\cdot\vec{\psi_0}(t_0) \psi_0(t_0+dt,x)+...
\end{align}
Note that we are ignoring the time dependence of $E_0$. It is safe to do so since $|E_0(t_0+dt)|>|E_0(t_0)|$, and so the growth in the ground state is actually underestimated in the above expression. When $t_0$ is very small, say, equal to $dt$, then the width of the well doubles in one time step. This means the inner product of the new and old ground states appearing in the coefficient is expected to be significantly smaller than one. However, as $t_0$ increases, the relative change in the ground state becomes smaller and smaller, and so the inner product approaches $1$. Simultaneously, the energy $E_0(t_0)$ approaches $V''_{2B-A}$. So, we argue that for any (negative) $V''_{2B-A}$, there is always a $t_0$ large enough that $E_0(t_0)$ is sufficiently negative that
\begin{equation}
1\leq \exp(\sqrt{|E_0(t_0)|}dt) \vec{\psi_0}(t_0+dt)\cdot\vec{\psi_0}(t_0)
\end{equation}
and the remaining contributions from components in other eigenstates are negligible (i.e. the ground state wavefunction at $t_0$ is "nearly" orthogonal to the other eigenstates at $t_0+dt$ because the new ground state wavefunction overlaps so much with the old ground state. The components in other eigenstates with negative energies will also of course have exponentially growing and decaying factors, but the ground state is the negative energy with the largest magnitude, so the ground state component dominates). Note that if $\dot\sigma(t_0,x)\ne0$ the argument still holds, there will simply be a term proportional to $dt$ in the following expression, but it will be dominated by the exponential term for $E_0(t_0)$ sufficiently negative.

\end{document}